\documentclass[conference]{IEEEtran}
\IEEEoverridecommandlockouts
\usepackage{cite}
\usepackage{amsmath,amssymb,amsfonts}
\usepackage{algorithmic}
\usepackage{graphicx}
\usepackage{subcaption}
\usepackage{textcomp}
\usepackage{xcolor}
\usepackage{comment}
\usepackage{hyperref}
\def\BibTeX{{\rm B\kern-.05em{\sc i\kern-.025em b}\kern-.08em
    T\kern-.1667em\lower.7ex\hbox{E}\kern-.125emX}}
\begin{document}

\title{Now You're Thinking With Structures: \\
	A Concept for Structure-based Interactions with Spreadsheets}
%\title{Refactoring Operations for Spreadsheet Structures
%*\\

\author{
	\IEEEauthorblockN{
		%1\textsuperscript{st} 
		Patrick Koch}
	\IEEEauthorblockA{
		%\textit{dept. name of organization (of Aff.)} \\
		\textit{AAU Klagenfurt}\\
		Klagenfurt, Austria \\
		Email: Patrick.Koch@aau.at}
	%\and
}

\maketitle

\begin{abstract}
Spreadsheets are the go-to tool for computerized calculation and modelling, but are hard to comprehend and adapt after reaching a certain complexity.
In general, cognition of %and deliberate intervention within 
complex systems is facilitated by having a higher order mental model of the system in question to work with.
We therefore present a concept for structure-aware understanding of and interaction with spreadsheets that extends previous work on structure inference in the domain.
Following this concept, structural information is used to enrich visualizations, reactively enhance traditional user actions, and provide tools to proactively alter the overall spreadsheet makeup instead of individual cells.
The intended systems should, in first approximation, not replace common spreadsheet tools, but provide an additional layer of functionality alongside the established interface.
In ongoing work, we therefore implemented a tool for structure inference and visualization along the common spreadsheet layout.
Based on this framework, we plan to introduce the envisioned proactive and reactive interaction mechanics, and finally provide structure-aware functionality as an add-in for common spreadsheet processors.
We believe that providing the tools for thinking about and interacting with spreadsheets in this manner will benefit users both in terms of productivity and overall spreadsheet quality.
\end{abstract}


\section{Introduction}
Tools for end-user programming, first of all spreadsheet processors like Microsoft Excel, enjoy great popularity with business and private users alike.
An analysis of a 2012 survey involving 5010 US citizens \cite{DBLP:conf/vl/Scaffidi17} concluded that 45\,\% of respondents used spreadsheets at work at least monthly, and workers who used spreadsheets earned more than their peers, all else being equal.
Indeed, the combination of tangible units of calculation in cells and rapid feedback due to instantaneous formula evaluation in spreadsheets provides a powerful interface for on-demand development of custom calculation models.
However, the same calculation-result-first paradigm that poses such an easy point of access to spreadsheets turns out to also be a major drawback in terms of understandability for advanced spreadsheet models \cite{DBLP:journals/ijmms/HendryG94}.
In cases where the functionality of a sheet is not immediately apparent by the displayed numbers and layout, further understanding often requires inspection of formulas and reference chains.
However, due to overly complex formulas and poor spreadsheet layouts, inferring the functionality of a spreadsheet by formula inspection can be challenging even for professional programmers.

To alleviate this issue, a number of different approaches for improving spreadsheet QA have been developed in recent years \cite{DBLP:journals/jss/JannachSHW14}. 
Continuing this intent, we propose to reinforce a more abstract mindset of spreadsheets, that focuses on overarching design and cohesion instead of functionalities of individual cells.
To support this perspective, we outline an interaction scheme for spreadsheets that expands on the functionality of conventional spreadsheet processors by means of information about higher-order structures that are already present in the sheets.
The required structural information is provided by an automated structure inference process that was presented in previous work \cite{DBLP:conf/issre/KochHW16}.
In particular, we propose to:
(1)~enhance spreadsheet visualization by adding structure information either in-place or via an additional, complementary UI element; 
(2)~include reactive procedures that trigger after common user actions and automatically restore structural soundness; and
(3)~provide a set of proactive methods for directly altering the overall calculation structure of a spreadsheet.
The basic ideas in this list are not new.
We do however believe that providing users with a concise combination of information and tools to interact with the overall structure of their spreadsheets will lead to more awareness for spreadsheet quality in general and will consequently result in more efficient and less error prone spreadsheets.
In ongoing work, we tested preliminary variants of different visualization schemes that follow the outlined guidelines in a research prototype\footnote{download via \url{http://spreadsheets.ist.tugraz.at/index.php/software/fritz/}}, and found that visualization of coherent cell groups can for example greatly assist in comprehension of spreadsheet models.

The remainder of the paper is structured as follows:
We first describe the fundamental structure primitives along a motivating example in Section~\ref{sec:example}.
The subsequent sections each consider one of the three suggested areas for improvement along the terms of \emph{Design Intent}, \emph{Technical Considerations}, an \emph{Example}, and \emph{Impact \& Discussion}:
Section~\ref{sec:visualization} suggests spreadsheet structure visualization approaches;
Section~\ref{sec:reactive} outlines a reactive procedure that monitors user actions in spreadsheets; and
Section~\ref{sec:proactive} discusses proactive spreadsheet structure interactions.
In Section~\ref{sec:related_work}, we briefly summarize previous works, and Section~\ref{sec:conclusion} concludes the work.

\section{Motivating Example} \label{sec:example}
%The structural components we want to highlight in this work are based on a previous approach for structure inference presented in \cite{DBLP:conf/issre/KochHW16}.
%The basic building blocks that were investigated in this study are \emph{Formula groups}, \emph{Reference groups}, and \emph{Calculation blocks}, which we will introduce along the following example.

\begin{figure}

	\centering
	\begin{subfigure}[b]{0.5\textwidth}
		\includegraphics[width=\textwidth]{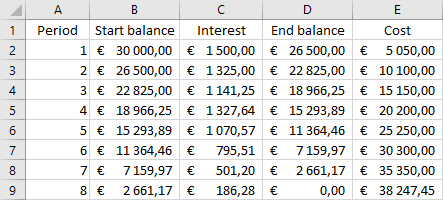}
		\caption{Value view.}
		\label{fig:value}
	\end{subfigure}
	%~ %add desired spacing between images, e. g. ~, \quad, \qquad, \hfill etc. 
	\par\medskip % force a bit of vertical whitespace
	\begin{subfigure}[b]{0.5\textwidth}
		\includegraphics[width=\textwidth]{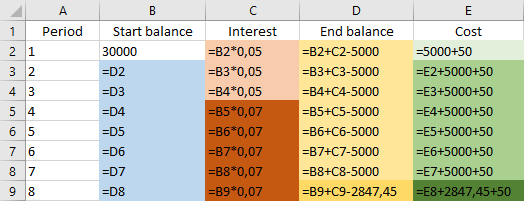}
		\caption{Formula view with highlighted formula groups.}
		\label{fig:formula_groups}
	\end{subfigure}
	%~ %add desired spacing between images, e. g. ~, \quad, \qquad, \hfill etc. 
	\par\medskip % force a bit of vertical whitespace
	\begin{subfigure}[b]{0.5\textwidth}
	\includegraphics[width=\textwidth]{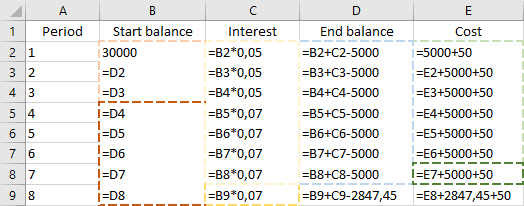}
	\caption{Formula view with highlighted reference groups.}
	\label{fig:reference_groups}
\end{subfigure}
	\setlength{\abovecaptionskip}{-3pt}
	\setlength{\belowcaptionskip}{-13pt}
	\caption{Example spreadsheet of a car loan.}\label{fig:example}
\end{figure}

Figure~\ref{fig:example} shows a car loan calculation spreadsheet.
Given the value view of the sheet only, as illustrated by Subfigure~\ref{fig:formula_groups}, the calculations of the example seem straightforward at first glance.
However, when investigating the example in greater detail, as provided by the formula view in Subfigure~\ref{fig:formula_groups}, it becomes apparent that the end result is depending on a number of interwoven and, most importantly, inconsistent calculations.

The highlighting of \emph{Formula groups} facilitates distinguishing between the different calculations that take place.
These groups summarise a number of neighboring cells that fulfil the same purpose in the sheet by applying the same type of calculation.
For example, the formula group \texttt{B3:B9} provides the \emph{Start balance} of the current year by referring to the \emph{End balance} of the previous year.
Each cell within a formula group might refer to other cells via a number of references.

\emph{Reference groups}, in turn, are defined as the set of cells that are referred to by the cells of a formula group via one specific reference of the group's formula.
Subfigure~\ref{fig:reference_groups} shows the reference groups that are inferred for the colour-linked formula groups in Subfigure~\ref{fig:formula_groups} using hatched borders.
For example, the formula group \texttt{D2:D8} of the \emph{End balance} column refers to the reference group \texttt{C2:C8} via the second reference (\texttt{$C_n$}) in the group's formula (\texttt{=$B_n$+$C_n$-5000}). Note that another reference group, \texttt{B2:B8}, is also inferred by the first reference of the group (\texttt{$B_n$}), but is not displayed in the example, as it coincides with the borders of other groups for the same cells.
This illustrates one major difference between formula groups and reference groups: each cell can only be part of one formula group, but can be part of multiple coinciding reference groups.
Moreover, reference groups can but do not necessarily have to coincide with formula groups that contain the same cells.

\begin{comment}
\section{Structure-based Interaction Mechanics}
So, how can we use structures to benefit the interaction with spreadsheets:
on one hand, we can visualize the structural make-up and connections within a given spreadsheet.
On the other hand, we can make use of structure information to either proactively or reactively enhance the actions with which a user influences a spreadsheet. 
\end{comment}

\section{Structure Visualization in Spreadsheets}\label{sec:visualization}
The following section addresses proposed enhancements of spreadsheet visualization schemes by addition of structure information.
\emph{Design Intent.}
The main goal of structure visualization approaches should be to point out relevant information about the positions, locations, and extents of the different structures that are present in a given spreadsheet.
The UI should also be able to adequately express the connections in-between those structures, and should also make apparent which further structure-based interactions (see Section \ref{sec:proactive}) are available, and what effects these interactions would have.

\emph{Technical Considerations.} 
In common spreadsheet UIs, feedback about membership of individual cells in any structure is likely best represented by either using matching cell background colours or cell borders to highlight cells.
Another approach would be to include some form of additional graphical interface component which provides contextual information either for the current sheet or the currently selected cell.
In either case, the amount of available data to display would make some form of information filtering necessary.
We therefore suggest to provide different, user-selectable \emph{perspectives}, that highlight a certain structural aspect (e.g. the formula groups) of either the current worksheet, or a specific selected cell.

%As pointed out above, the information we need to visualize is not limited to membership of cells in structures, but also encompasses relational information like connections to other cells and worksheets via references, or connections to related header cells.

\emph{Example.}
Figures~\ref{fig:formula_groups} and \ref{fig:reference_groups} provide examples for simple but effective visualizations. 
Background colour and borders are used as stylistic devices to point out membership of different cells in either formula groups or reference groups.
%Our research prototype implementation provides similar visualization schemes that are selectable via a drop-down UI element. 

Spreadsheet structures, e.g. formula groups, could also be represented in a supplemental graph layout, that connects the different groups based on inter-group references. The UI could then navigate to and highlight a group or connection in the common UI, that was selected within the graph visualization.

\emph{Impact \& Discussion.}
Visual representations are the primary means of relaying information to the users of software systems.
Therefore, adequate visual emphasis of the different inherent structural properties of a given spreadsheet is the first and foremost directive for raising awareness and support for structural aspects of spreadsheet design.

The main concern for the implementation of such a visualization scheme would likely be not to overwhelm users with the new and unexpected information.
Ideally, this information would seamlessly blend in the already familiar spreadsheet UI.
As the available information is likely to exceed the available real estate for tactful integration in existing graphical interfaces, some form of guided user interaction to select the currently relevant information would be beneficial.
In addition, spreadsheet processors could provide some form of user-initiated process that checks for structural issues, and highlights individual issues along with a matching visualization. 

\section{Structure-preserving Procedures} \label{sec:reactive}
The following section discusses reactive procedures that automatically restore structural soundness.
\emph{Design Intent.}
The intent of this core theme is to provide some form of process that, after each user action that altered a given spreadsheet, checks whether the overall structure of the sheet is still sound.
Otherwise, the process suggests some form of additional action to performed which restores structural soundness.
The system should provide appropriate feedback based on the analysis result, along with the possibility to fix structural issues automatically when possible, and to give suggestions about necessary manual interventions when not.

\emph{Technical Considerations.} 
The first obvious consideration regards the functional requirements to a system that supports the envisioned checks and alterations.
Such a system would have to either preprocess and store an abstract representation of the structural makeup of a sheet for each cell, or be able to deduce the necessary structural relations for any given alteration in real time.
It would also require some defined notion of structural soundness that is expressible for the inferred structures, and it would need the capability to check for violations of soundness, as well as to automatically generate and apply candidates for repair operations.
In many cases, e.g. the expansion of a formula to include a new reference to an hitherto empty cell, the system would also be required to be able to prompt the user for some form of input.

The second consideration regards the question as to how such a system could be integrated into common spreadsheet tools.
We assume that any new interaction affects a spreadsheet that was structurally well-formed before the interaction (this could be established in an initial analysis and repair step).
In that a case, the addition or deletion of cells that are part of existing formula groups or reference groups can automatically be cascaded by addition or removal of similar cells within other, functionally related cell groups.
In cases where user input is required, the intended alteration could either be displayed by highlighting the affected cell(s), and either prompting textual input, or providing predefined suggestions in form of a drop-down list.
In general, the feature should remain optional, allowing users to deactivate the functionality, if no automatic checking and alteration is requested.

\emph{Example.}
Every action that directly influences the overall structure of a spreadsheet would require a subsequent soundness check. 
This pertains in particular actions that add, delete, or alter the content of an individual cell.
Actions that affect multiple cells at once can be assumed to be consecutive alterations of individual cells.
In our running example in Figure~\ref{fig:example}, the user could alter the formula in cell \texttt{C6} to apply another interest rate. 
A suggested reaction would then be to apply the same interest rate for the remainder of the group \texttt{C5:C9}.
Deletion of the cell \texttt{C6} would be more challenging:
as the cells are connected by an ongoing chain of references, all subsequent calculations would normally be missing a reference.
However, using structural information, we can deduce that the respective cells in the other formula groups in Row 6 can also be matched to the same calculation chain, and therefore can be removed as well. 
This allows us to reconnect the remaining cell references accordingly. 
However, in many cases such alterations can be fixed by a number of repair candidates, which makes user involvement necessary.

\emph{Impact \& Discussion.}
When using spreadsheet systems that check and establish a form of structural soundness, we could guarantee the absence of specific fault types (e.g. typos) in the created spreadsheet models.
However, most users are likely overwhelmed when being confronted unexpectedly with the such a feature.
Some form of guided introduction for the process and its notion of soundness would be required.

Moreover, following such an approach also severely limits the design freedom of the spreadsheet paradigm.
Especially for the development of small prototype sheets or specialized calculations, the constant interference of the corrective script would pose a hinderance, and should therefore be deactivatable.
In practice, some form of guided transition mechanism would be desirable, which takes a spreadsheet that was developed free-form as input, and converts it to a structurally sound version that provides the same functionality.

\section{Structure-altering Operations} \label{sec:proactive}
The following section discusses methods for altering the overall structure of a spreadsheet.
\emph{Design Intent.}
This last proposed enhancement for spreadsheet interfaces aims at providing tools to deliberately alter the overall makeup of their spreadsheets, that go beyond alteration of individual cells.
Such operations allow users to add, alter, relocate, and remove entire structural components of a given sheet, while keeping its layout and calculation integrity intact. 
Such tools are, however, closely dependent on an adequate visual representation that points out such options for interaction and also allows for assessment of possible consequences (see Section~\ref{sec:visualization}).

\emph{Technical Considerations.} 
Technically, structure-altering operations can, in simple cases, be lead back to incremental application of atomic changes and handling the cascading automatic adaptations accordingly (see Section~\ref{sec:reactive}).
In general, however, these operations represent deliberate changes in the overall structure of a sheet, and we thus can make use of all structural information to plan the actions beforehand.  

\emph{Example.}
An example for such an operation could be splitting up a formula group into two sets of linked calculations: we fist determine which external reference groups are referred to by the group in question, and which other formula groups refer to it.
Based on this information, we can then introduce a new space for the second formula group next to their existing position, initializing it with a copy of the base group.
Next, we decide at which point to split the base formula, and alter the formulas of the two groups accordingly.
Lastly, we re-arrange the formula-references of all participating formula groups based on the new layout and calculation order.

\emph{Impact \& Discussion.}
The proposed tools can be interpreted as refactoring operations for spreadsheets that operate on a global, structural scope.
Using these operations, specific shortcomings and issues in the overall composition of a spreadsheet can be handled with relative ease.
Such issues are indeed relatively common, and can be detected in the form of spreadsheet smells \cite{DBLP:conf/icsm/HermansPD12, DBLP:conf/icse/HermansPD12, DBLP:conf/iccsa/CunhaFRS12}.
In general, providing these tools invites users to preserve an overall good spreadsheet quality, as individual maintenance actions can be performed without fear of unexpected consequences due to overlooked dependencies.
This, in turn, raises awareness for good spreadsheet quality, which is likely to reflect also in better constructed spreadsheets in the future, that are less susceptible to errors.

\section{Related Work} \label{sec:related_work}

In terms of previous works that analyzed structural properties of spreadsheets, Hermans \textit{et~al.} extracted class diagrams and dataflow diagrams from spreadsheets \cite{DBLP:conf/ecoop/HermansPD10, DBLP:journals/corr/abs-1111-6895}, Abraham and Erwig inferred areas of calculation and related headers \cite{DBLP:journals/vlc/AbrahamE07}, Mittermeir and Clermont inferred logical areas and semantic classes from spreadsheets \cite{DBLP:conf/wcre/MittermeirC02}, Kankuzi and Ayalev worked on a graph-based visualizations of spreadsheets based on Markov Clustering (MCL) \cite{kankuzi2008end}, and Schmitz and Jannach provide structure-considering tool support for spreadsheet debugging \cite{DBLP:conf/vl/SchmitzJ17}.
In our own research, we inferred groups, blocks, and related headers for spreadsheets \cite{DBLP:conf/issre/KochHW16}.

Abraham end Erwig introduced formal descriptions for spreadsheet templates \cite{DBLP:conf/icse/AbrahamE06}.
Cunha \textit{et~al.} adapted the idea, provided approaches to define and infer formal spreadsheet models, and also proposed safe spreadsheet operations and refactorings based on inferred model information \cite{DBLP:conf/vl/CunhaSV09, DBLP:conf/vl/CunhaES10,DBLP:journals/tse/CunhaFMS15, DBLP:journals/ase/CunhaEMS16, DBLP:journals/jss/CunhaFMMPS16}.

%Hermans \textit{et~al.} \cite{DBLP:conf/icsm/HermansPD12, DBLP:conf/icse/HermansPD12} established the notion of spreadsheet smells to point out structural deficits in spreadsheets, that was expanded on by Cunha \textit{et~al.}~\cite{DBLP:conf/iccsa/CunhaFRS12}.

Previous research on impact aware editing and refactoring in spreadsheets was conducted by Hermans \textit{et~al.}~\cite{DBLP:conf/sigsoft/HermansD14, DBLP:journals/ese/HermansPD15}, Badame and Dig~\cite{DBLP:conf/icsm/BadameD12}, Cunha \textit{et~al.}~\cite{DBLP:conf/vl/CunhaSV09}, and O'Beirne \cite{DBLP:journals/corr/abs-1009-1412}.

\section{Conclusions \& Outlook} \label{sec:conclusion}

We presented a three-pronged conceptual approach for extending interactions with spreadsheet systems by means of structural information.
Major challenges that we identified are the filtering of appropriate information as to not overwhelm users, and the conceptualization of adequate user interaction methods where user input is required.
Nevertheless, we believe that providing structure-enhanced information and tools can benefit the way users handle and think about spreadsheets.

In intermediate future work, we plan to further investigate structure visualization approaches and structure-aware refactoring operations for spreadsheets.
Furthermore, thinking about spreadsheet in structures opens up a number of interesting further potential applications.
For example, machine-learning approaches can exploit inferred structural representations to learn which features are conductive for good spreadsheet design, and consequently offer automated design feedback or even spreadsheet synthesis.
Another interesting application could be the inference of topic and contextual information, by linking the extracted structures with semantic databases.
Looking beyond the spreadsheet domain, raising awareness for structural connections in general helps users to better conceptualize and work with computational systems.

\section*{Acknowledgment}
The work described in this paper has been been funded by the Austrian Science Fund (FWF) project {\em DEbugging Of Spreadsheet programs (DEOS)} under contract number I2144.
%\section*{References}
\bibliographystyle{IEEEtran}
%\bibliography{IEEEabrv,references}


\end{document}